\documentclass[conference]{IEEEtran}
\IEEEoverridecommandlockouts 
\usepackage[utf8]{inputenc}
\usepackage{orcidlink}
\usepackage{cite}

\usepackage{xcolor}
\usepackage{colortbl}

\usepackage{amsmath,amssymb,amsfonts}
\usepackage{algorithmic}
\usepackage{graphicx}
\usepackage{textcomp}

\definecolor{qnowcolor}{HTML}{364483}
\usepackage{hyperref}
\hypersetup{
  colorlinks=true,
  allcolors=qnowcolor
}
\usepackage{booktabs}
\usepackage{hhline}
\usepackage{array, makecell}
\usepackage{mathrsfs}
\usepackage[scr=dutchcal,calscaled=1]{mathalfa}
\usepackage{braket}
\usepackage{pifont}

\usepackage{url}

\usepackage{comment}

\def\BibTeX{{\rm B\kern-.05em{\sc i\kern-.025em b}\kern-.08em
    T\kern-.1667em\lower.7ex\hbox{E}\kern-.125emX}}

\begin{document}

\title{$\mathtt{Q^2SAR}$: A Quantum Multiple Kernel Learning Approach for Drug Discovery}

\author{
\IEEEauthorblockN{Alejandro Giraldo \orcidlink{0009-0008-9826-0703}}
\IEEEauthorblockA{
\href{https://qnow.tech/}{$\mathtt{QNOW \; Technologies}$}
\\ Delaware, USA\\
\href{mailto:alejandro@qnow.tech}{alejandro@qnow.tech}}

\and

\IEEEauthorblockN{Daniel Ruiz \orcidlink{0009-0007-6976-1755}}
\IEEEauthorblockA{\href{https://qnow.tech/}{$\mathtt{QNOW \; Technologies}$}
\\ Delaware, USA
\\ \href{mailto:daniel@qnow.tech}{daniel@qnow.tech}
}

\and

\IEEEauthorblockN{Mariano Caruso  \orcidlink{0000-0002-7455-1193}}
\IEEEauthorblockA{
\href{https://www.ugr.es/}{$\mathtt{UGR}$}, Granada, Spain\\
\href{https://www.unir.net/}{$\mathtt{UNIR}$},
La Rioja, Spain\\
\href{https://www.fidesol.org/}{$\mathtt{FIDESOL}$}, Granada, Spain \\
\href{mailto:mcaruso@fidesol.org}{mcaruso@fidesol.org}
}

\and

\IEEEauthorblockN{Javier Mancilla \orcidlink{0000-0002-2862-9877}}
\IEEEauthorblockA{
\href{https://www.falcondale.pro/}{$\mathtt{Falcondale \; LLC}$}
\\ Delaware, USA\\
\href{mailto:javier@falcondale.pro}{javier@falcondale.pro}
}

\and

\IEEEauthorblockN{Guido Bellomo \orcidlink{0000-0001-8213-8270}}
\IEEEauthorblockA{
\texttt{CONICET} - UBA 
\\ 
\texttt{ICC}, Argentina\\
\href{mailto:gbellomo@icc.fcen.uba.ar}{gbellomo@icc.fcen.uba.ar}
}
}
\maketitle

\begin{abstract}
Quantitative Structure-Activity Relationship ($\mathtt{QSAR}$) modeling is a cornerstone of computational drug discovery. This research demonstrates the successful application of a Quantum Multiple Kernel Learning ($\mathtt{QMKL}$) framework to enhance $\mathtt{QSAR}$ classification, showing a notable performance improvement over classical methods. We apply this methodology to a dataset for identifying $\mathtt{DYRK1A}$ kinase inhibitors. The workflow involves converting $\mathtt{SMILES}$ representations into numerical molecular descriptors, reducing dimensionality via Principal Component Analysis ($\mathtt{PCA}$), and employing a Support Vector Machine ($\mathtt{SVM}$) trained on an optimized combination of multiple quantum and classical kernels. By benchmarking the $\mathtt{QMKL}$-$\mathtt{SVM}$ against a classical Gradient Boosting model, we show that the quantum-enhanced approach achieves a superior $\mathtt{AUC}$ score, highlighting its potential to provide a quantum advantage in challenging cheminformatics classification tasks.
\end{abstract}


\section{Introduction}
Drug discovery is an inherently complex and resource-intensive process. Computational methodologies, particularly Quantitative Structure-Activity Relationship ($\mathtt{QSAR}$) modeling, are essential for the efficient evaluation and optimization of chemical compounds \cite{Hansch1964}. $\mathtt{QSAR}$ models establish a mathematical link between a molecule's structure and its biological activity, enabling the \textit{in silico} prediction of properties for novel compounds, thereby prioritizing experimental efforts and reducing costs \cite{NatarajanEtAl2025Molecules}.

The evolution of $\mathtt{QSAR}$ has seen a shift from simple linear regression to sophisticated machine learning algorithms like Random Forests \cite{Svetnik2003} and Support Vector Machines ($\mathtt{SVMs}$) \cite{cortes1995support}, which excel at capturing complex, non-linear relationships. However, the continuous growth of chemical databases like ChEMBL \cite{gaulton2023chembl} means that molecular data is increasing in both volume and complexity. This high-dimensional feature space presents a fundamental challenge even for advanced classical algorithms.

Quantum Machine Learning ($\mathtt{QML}$) offers a promising new frontier to address this challenge. By mapping classical data into the exponentially large Hilbert spaces of quantum systems, $\mathtt{QML}$ algorithms have the potential to identify patterns intractable for classical models \cite{SchuldKilloran2022PRXQ, HavlicekEtAl2019Nature}. This work focuses on Quantum Multiple Kernel Learning ($\mathtt{QMKL}$) \cite{miyabe2023quantummultiplekernellearning}, an advanced technique that combines the strengths of various quantum and classical kernels to build a more robust and accurate classifier. We apply this framework to a real-world drug discovery use case: identifying inhibitors for the $\mathtt{DYRK1A}$ kinase, a critical therapeutic target for neurodegenerative disorders \cite{johnson2004tau}.

This paper details our complete workflow, from data curation and descriptor calculation to the implementation of the $\mathtt{QMKL-SVM}$ model. We provide a fair benchmark against a classical Gradient Boosting model and discuss the implications of our results, including the computational costs and future potential of this quantum-hybrid approach.

\section{Background and Related Work}

\subsection{Foundations of Quantitative Structure-Activity Relationships}
$\mathtt{QSAR}$ modeling is a cornerstone methodology in computational chemistry and drug discovery. The theoretical foundation rests on the premise that similar molecular structures exhibit similar biological activities, a principle that enables the systematic prediction of pharmacological, toxicological, and physicochemical properties from structural information alone \cite{Neubig2003}.

A typical $\mathtt{QSAR}$ workflow involves representing molecules (e.g., via $\mathtt{SMILES}$ strings), calculating numerical molecular descriptors, developing a model using machine learning, and validating its predictive performance \cite{carpio2021computational, tropsha2010best}. These descriptors can range from simple atom counts and topological indices to complex 3D features that capture the spatial arrangement of a molecule \cite{serrano_candelas2023}.

\subsection{Contemporary Relevance and Industrial Impact}
The relevance of $\mathtt{QSAR}$ has intensified due to the exponential growth of chemical data. $\mathtt{QSAR}$ offers three fundamental advantages in modern drug development:
\begin{itemize}
    \item \textbf{Predictive Efficiency:} Once validated, models allow for instantaneous property prediction, accelerating the assessment of large compound libraries.
    \item \textbf{Scalability:} Modern frameworks support the automated, high-throughput computational screening of millions of virtual compounds.
    \item \textbf{Regulatory Acceptance:} $\mathtt{QSAR}$ models are accepted by regulatory agencies like the $\mathtt{OECD}$, $\mathtt{EPA}$, and $\mathtt{EMA}$ for toxicity and risk assessment, providing a path from computation to clinical application.
\end{itemize}

\subsection{Quantum Frontiers in Cheminformatics}
Despite advances, classical machine learning methods face limitations in exploring high-dimensional feature spaces. The emergence of $\mathtt{QML}$ in cheminformatics represents a paradigm shift, aiming to harness quantum principles for enhanced molecular property prediction \cite{SuzukiEtAl2023Arxiv}. By mapping classical molecular data into quantum states, quantum algorithms can potentially access richer representational spaces and capture subtle structure-activity relationships that classical methods might overlook \cite{ChenEtAl2023Arxiv}.

\section{Methodology}
Our pipeline consists of three main stages: \textbf{(1)} data curation and feature engineering, \textbf{(2)} classification using a $\mathtt{QMKL}$-$\mathtt{SVM}$ model, and \textbf{(3)} benchmarking against a classical Gradient Boosting model. Predictive performance is assessed using standard metrics, with a focus on the Area Under the Receiver Operating Characteristic Curve ($\mathtt{ROC-AUC}$).

\subsection{Dataset Curation and Preprocessing}
This study is based on a high-quality dataset of $\mathtt{DYRK1A}$ inhibitors, provided and curated by ProtoQSAR SL from the ChEMBL database \cite{gaulton2023chembl}. $\mathtt{DYRK1A}$ is a critical therapeutic target for neurodegenerative disorders like Alzheimer's disease \cite{johnson2004tau}.

The raw data for 1291 molecules underwent rigorous curation: duplicates were removed, chemical structures were validated, and only assays labeled as "single protein assay" were retained to ensure direct target-compound interaction measurements. The final dataset comprised 354 unique molecules. For binary classification, a standard medicinal chemistry threshold was used: molecules with $pIC_{50} \geq 6.0$ (corresponding to $IC_{50} \leq 1~\mu M$) were labeled "active," and all others were labeled "inactive." The dataset was pre-divided into a training set (283 molecules) and a test set (71 molecules).

\subsubsection{Molecular Descriptors}
A comprehensive set of molecular descriptors was calculated using $\mathtt{RDKit}$ and validated to ensure they fell within typical drug-like ranges. Key descriptors included:
\begin{itemize}
    \item \textbf{Shape/Geometric:} Principal Moments of Inertia (e.g., \textbf{PMI1}) and Normalized Principal Ratios (e.g., \textbf{NPR1}).
    \item \textbf{Electronic/Topological:} Molecular LogP (\textbf{MolLogP}) and Topological Polar Surface Area (\textbf{TPSA}), where values $< 140$ \AA$^2$ often correlate with good blood-brain barrier permeability.
    \item \textbf{Connectivity:} Bertz Complexity Index (\textbf{BertzCT}) \cite{Bertz1981} and Kier-Hall Connectivity Indices (e.g., \textbf{Chi0n}).
    \item \textbf{Pharmacophore/Functional Group:} Counts of Hydrogen Bond Donors/Acceptors (\textbf{NumHBD}, \textbf{NumHBA}) and the fraction of sp$^3$ hybridized carbons (\textbf{FractionCsp3}).
\end{itemize}
This high-dimensional descriptor matrix provides a comprehensive mathematical representation of each molecule's structure.

\subsubsection{Data Processing Pipeline}
A multi-step preprocessing pipeline was implemented:
\begin{enumerate}
    \item \textbf{Numerical Feature Extraction}: Each $\mathtt{SMILES}$ string was converted into a vector of ~200 physicochemical descriptors using $\mathtt{RDKit}$.
    \item \textbf{Standardization}: Features were scaled to have a mean of 0 and a standard deviation of 1 using Scikit-learn's \texttt{StandardScaler}, fitted only on training data.
    \item \textbf{Dimensionality Reduction}: To prepare the data for quantum simulation, we applied Principal Component Analysis ($\mathtt{PCA}$). This aggressive reduction to four components, which capture 37.1\% of the total variance, was necessary to ensure compatibility with quantum simulators, despite the substantial loss of information from the original descriptor space. These four components serve as the final input features for all models.
\end{enumerate}

\subsection{Quantum Multiple Kernel Learning}
$\mathtt{SVMs}$ use a kernel function, $K(\mathbf{x}_i, \mathbf{x}_j)$, to compute similarity in a high-dimensional feature space. Our approach extends this with $\mathtt{QMKL}$.

\subsubsection{Quantum Kernels}
A quantum kernel is created by encoding classical data $\mathbf{x}$ into quantum states $\ket{\psi(\mathbf{x})}$ via a parameterized quantum circuit (a feature map) $U(\mathbf{x})$ \cite{RebentrostEtAl2014PRL}. The kernel value is the fidelity between the states:
\begin{equation}
    \mathcal{K}(\mathbf{x}_i, \mathbf{x}_j) = \left| \braket{\psi(\mathbf{x}_i) | \psi(\mathbf{x}_j)} \right|^2 = \left| \bra{0} U^\dagger(\mathbf{x}_i) U(\mathbf{x}_j) \ket{0} \right|^2.
\end{equation}
This process can map data into a richer feature space \cite{HuangEtAl2021NatComm}. We used a feature map of angle-embedding rotations and $\mathtt{CNOT}$ entangling gates, implemented in PennyLane \cite{bergholm2018pennylane}.

\subsubsection{Multiple Kernel Learning and Weight Optimization}
Rather than relying on a single kernel, $\mathtt{MKL}$ seeks to find the best linear combination of a pre-defined set of basis kernels $\{K_m\}$:
\begin{equation}
    K_{comb} = \sum_{m=1}^M w_m K_m, \quad \text{with} \quad w_m \ge 0, \quad \sum_{m=1}^M w_m = 1.
\end{equation}
Our basis set included 8 distinct quantum fidelity kernels (from varied feature maps) and 3 classical Radial Basis Function (RBF) kernels. To determine the optimal weights $w_m$, we implemented and compared multiple strategies within our framework. The two primary methods evaluated were: 1) a simple Averaging approach, where each of the $M$ kernels is given an equal weight ($w_m = 1/M$), and 2) a selection strategy based on Centered Kernel Alignment (CKA). The CKA method measures the similarity between each base kernel $K_m$ and an ideal target kernel $K_y = yy^T$ (where $y$ is the vector of class labels), and our implementation selects the single kernel with the highest alignment score.

Upon evaluation, the CKA selection strategy consistently yielded the best-performing model. Therefore, the results reported in this paper correspond to the SVM trained on the single, best-aligned kernel identified through this process.

\subsection{Classical Benchmark Model}
A classical \texttt{GradientBoostingClassifier} from Scikit-learn was trained and evaluated on the exact same 4-dimensional, $\mathtt{PCA}$-processed data to ensure a fair comparison.

\section{Results and Discussion}
The performance of the final $\mathtt{QMKL-SVM}$ model and the classical Gradient Boosting model was evaluated on the unseen test set. Table \ref{tab:full_results} summarizes the performance across multiple metrics.

Our proposed $\mathtt{Q^2SAR}$ model showed superior overall performance, outperforming the baseline in the key metrics of AUC, Accuracy, Recall, and F1-Score. While the classical model achieved slightly higher precision (0.8333 vs. 0.8070), the significant gain in recall (sensitivity) from the quantum-enhanced model (0.9200 vs. 0.8000) indicates a much stronger ability to correctly identify true active compounds, which is often a primary goal in initial drug screening campaigns.

\begin{table}[htbp]
\centering
\caption{Comprehensive Performance Comparison on the DYRK1A Test Set}
\label{tab:full_results}
\begin{tabular}{lcc}
\toprule
\textbf{Metric} & \textbf{Gradient Boosting} & \textbf{$\mathtt{Q^2SAR}$ Model} \\
\midrule
\rowcolor{qnowcolor!10}
\textbf{AUC} & 0.8037 & \textbf{0.8750} \\
Accuracy & 0.8000 & \textbf{0.8333} \\
Precision & \textbf{0.8333} & 0.8070 \\
Recall (Sensitivity) & 0.8000 & \textbf{0.9200} \\
F1-Score & 0.8163 & \textbf{0.8598} \\
\bottomrule
\end{tabular}
\end{table}

\begin{figure}[htbp]
\centerline{\includegraphics[width=0.9\columnwidth]{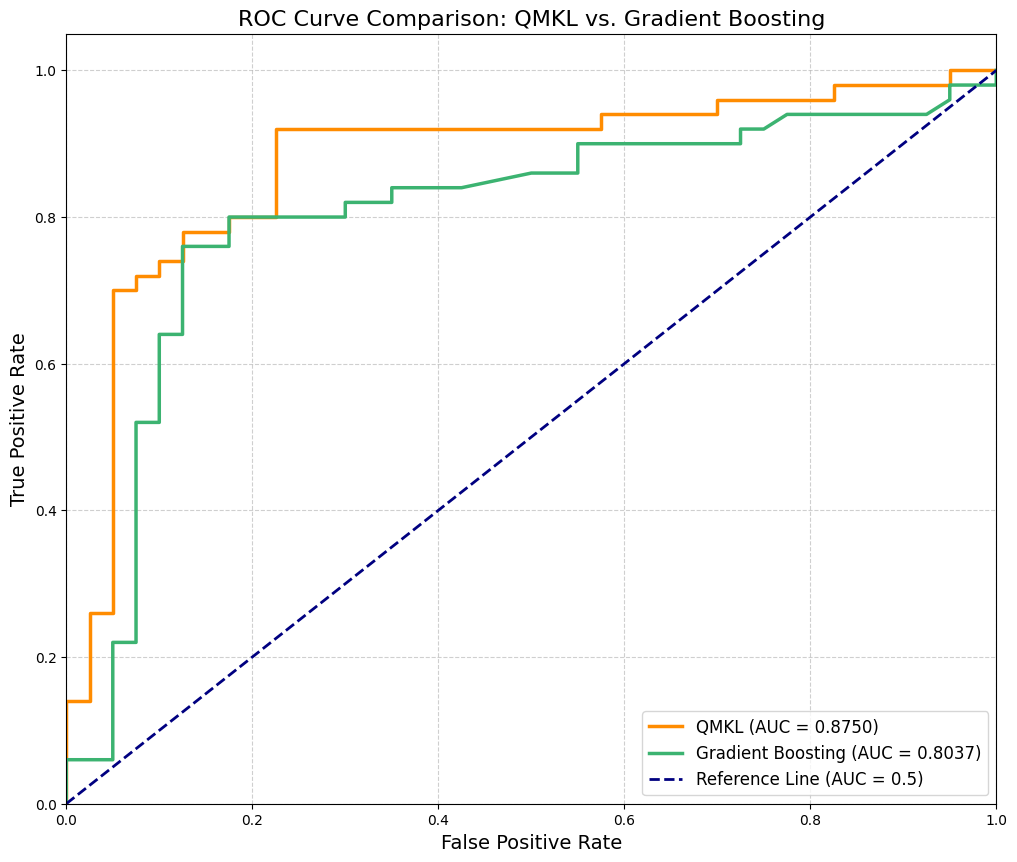}}
\caption{Comparison of ROC curves for the $\mathtt{QMKL-SVM}$ and Gradient Boosting models on the $\mathtt{DYRK1A}$ test set, with performance metrics detailed in Table \ref{tab:full_results}.}
\label{fig:roc_curve}
\end{figure}

Figure \ref{fig:roc_curve} provides a visual comparison of the $\mathtt{ROC}$ curves. The plot clearly illustrates the superior discriminative power of the $\mathtt{QMKL-SVM}$, as its curve consistently stays above the Gradient Boosting curve, encompassing a larger area. The results suggest that the optimally chosen kernel was able to capture complex, non-linear relationships within the data that were advantageous for identifying active molecules, even at the expense of slightly lower precision.

\section{Computational Cost and Scalability}
A fundamental challenge for kernel methods is scalability. Training a conventional SVM scales polynomially (typically $O(n^2)$ to $O(n^3)$) with the number of samples $n$, due to the need to compute the $n \times n$ kernel matrix. For quantum kernels, this is exacerbated, as building the matrix requires $O(n^2)$ quantum circuit executions, a significant bottleneck for near-term quantum hardware. It is critical to note that all experiments in this study were performed on a classical simulator. On real quantum devices, challenges such as hardware noise and decoherence would further impact the fidelity of the kernel computation and must be addressed with error mitigation techniques.

Strategies to mitigate this exist, such as quantum kernel estimation, using representative data centroids (e.g., $\mathtt{QuACK}$ \cite{Tscharke2024}), or reformulating the SVM optimization to scale linearly (e.g., PegasosQSVC \cite{ShalevShwartz2007}). These approaches will be crucial for achieving practical quantum advantage on large-scale problems.

\section{Conclusion}
In this work, we developed a complete $\mathtt{QSAR}$ pipeline demonstrating that a quantum-enhanced model can outperform a strong classical baseline. Our $\mathtt{QMKL-SVM}$ approach achieved a superior $\mathtt{AUC}$ score (0.875 vs. 0.804) on the $\mathtt{DYRK1A}$ inhibitor dataset, highlighting the potential of quantum feature spaces in drug discovery. The model's high recall, in particular, suggests its utility for hit-finding in early-stage discovery, where minimizing false negatives is paramount.

We acknowledge the limitations of this study, including the reliance on a single train-test split and the inherent "black-box" nature of kernel models. Future work should employ more rigorous validation schemes like cross-validation and explore model interpretability with techniques such as Kernel-SHAP to understand which chemical features drive the predictions.

As quantum hardware matures, overcoming the challenges of noise and scalability will be paramount. When that happens, frameworks like $\mathtt{Q^2SAR}$ could become impactful tools for accelerating the discovery of new medicines.

\section*{Acknowledgment}
This work was supported by the project $\mathtt{ECO-20241014}$ $\textcolor{qnowcolor}{\mathtt{QUORUM}}$ funded by the Ministerio de Ciencia, Innovación y Universidades, through $\mathtt{CDTI}$. We gratefully acknowledge ProtoQSAR SL and the team led by Eva Serrano-Candelas, Laureano E. Carpio, and Rafael Gozalbes for their fundamental contribution in providing, curating, and validating the dataset used in this study. For more information, refer to \textit{Computational Modeling of DYRK1A Inhibitors as Potential Anti-Alzheimer Agents} \cite{serrano_candelas2023}.

\bibliographystyle{IEEEtran}

\bibliography{IEEEabrv, references}

\end{document}